# The interaction between phosphorene oxide and the villin headpiece

Wei Zhang[1,*], Yuanyuan Gou[1], Li Cheng[1], Chao Ye[2] and Xianqing Yang[1,*]

[1]School of Materials and Physics, China University of Mining and Technology, Xuzhou 221116, China
[2]Wenzhou University of Technology, Wenzhou 325000, China

**ABSTRACT:**
Phosphorene, a novel member of the two-dimensional nanomaterial family, has been demonstrated a great potential in biomedical applications, such as photothermal therapy, drug delivery and antibacterial. However, phosphorene is unstable and easily oxidized in an aerobic environment. In this paper, using the larger-scale molecular dynamics simulations, we investigated the disruption of phosphorene oxide (PO) to the structure of a model protein, villin headpiece subdomain (HP35). It shows that the disruption of PO nanosheets to protein structure enhances with the increase of the oxidation concentration of PO, but the oxidation mode has almost no effect on the PO-HP35 interaction. At low oxidation concentration, PO is good biocompatibility to HP35. Oxygen atoms filled into the groove region in puckered surface of phosphorene enhances the dispersion interaction between phosphorene and HP35. Thus, oxidation enhances the destructive effect of phosphorene on the structure of HP35. These findings might shed light on understanding the biological toxicity of PO nanosheets and would be helpful for the future potential biomedical applications of PO nanosheets, such as nanodrugs and antibacterial agents.

**KEYWORDS:** phosphorene oxide; HP35; protein; biological toxicity

## 1. Introduction

In recent years, two-dimensional (2D) nanomaterials such as graphene, borophene, MoS2, hexagonal boron nitride (h-BN), transition metal dichalcogenides (TMDs), etc.

have attracted widespread attention due to their excellent electrical, mechanical, optical and chemical properties [1-3]. Most 2D nanomaterials have been used in the field of biomedicine [4-7]. Phosphorene is a novel member of 2D nanomaterials, with high charge-carrier mobility, adjustable direct bandgap, anisotropy, etc. [8-13] These characteristics make phosphorene superior to other 2D nanomaterials, such as graphene, silicon and MoS2, in terms of optical and optoelectronic properties. Phosphorene has attracted great attention in the fields of nanoelectronics and optoelectronics after it was successfully fabricated [14-18]. Phosphorene's unique topological structure and optoelectronics characteristics make it an ideal candidate for local biodetector, biosensor and drug delivery [19-21].

Recent experimental results have confirmed that black phosphorus composed of several layers of phosphorene is excellent in killing cancer cells [22-26], which opens a window for the application of phosphorene in the field of biomedicine. At the same time, the biological safety (toxicity) of phosphorene has also begun to attract people's attention [27-29]. Pumera et al. believe that the cytotoxicity of layered black phosphorus is between graphene oxide (GO) and transition metal chalcogenide TMD (MoS2, WS2, WSe2) [27]. The experimental results of Jouhahn Lee's group showed that black phosphorus has no or minimal cytotoxicity [28]. Molecular dynamics simulation results show that compared with graphene, phosphorene has less damage to the protein structure, but it is still biologically toxic [30]. Phosphorene could snatch the ligand [31], block signal pathways [32], and destructively extract phospholipid molecules from the membrane [33]. At present, research on the biological toxicity of phosphorene is in its infancy, and related experimental work is at the cellular level. There is still a lack of research on the molecular scale to analyze the toxicity mechanism of phosphorene.

Phosphorene is unstable in an aerobic environment and is easily oxidized [34]. Compared with the pristine phosphorene, the optoelectronic and mechanical characteristics of phosphorene oxide (PO) have changed [35, 36]. At present, some scholars suggested that oxygen, water and light are necessary conditions for PO [37], while others argued water is not necessary and phosphorene can be oxidized with oxygen and light [38]. The oxidation rate of phosphorene is proportional to the oxygen concentration and light intensity. Water plays an important role in the oxidation of phosphorene. Phosphorene is stable in water without oxygen (far more stable than in air) [23], but in water with sufficient oxygen, water will accelerate the oxidation of phosphorene [39, 40]. The stability of phosphorene in the water creates opportunities for the application of phosphorene in the field of biomedicine, and also provides a reference for people to control and eliminate the oxidation of phosphorene.

The first-principles calculation results show that there are three typical phosphorene oxidation modes: dangling oxidation, interstitial oxidation, horizontal bridge oxidation [38]. The calculation results of binding energy show that dangling and interstitial oxidation are the two oxidation modes that are easier to achieve, and horizontal bridge oxidation can be realized only by increasing the light energy. After oxidation, the surface morphology and curvature of the phosphorene altered significantly [41]. Due to the difference in electronegativity of phosphorus and oxygen atoms, electrical polarizations with different degrees appear on the surface of PO nanosheet. Such changes in surface electrical polarization and morphology would affect the hydrophilicity and hydrophobicity of phosphorene nanosheet, and would also have an important impact on the interaction between PO nanosheet and biomolecules. Phosphorene oxide nanosheets destructively extract phospholipid molecules on the phospholipid bilayer membrane and destroy the membrane structure [42]. Would PO nanosheets destroy the structure of proteins and cause toxicity? At present, there is no related report on the interaction between PO nanosheet and protein molecules, and a systematic study is needed.

In this paper, using large scale molecular dynamics (MD) simulation methods, we systematically investigate the interaction between PO and model protein HP35. We analyzed the effect of the oxidation mode and concentration of phosphorene on the interaction of PO-HP35, and presented the dynamic process of protein structure on the surface of PO. We observed that oxidation enhances the disruption of phosphorene nanosheet to protein structure. These findings can help one understand the biological toxicity of PO nanosheet at the molecular level, and can also provide a theoretical reference for scholars to design a new type of phosphorene-based biosensor.

## 2. Methods

To decipher the interactions between a PO nanosheet and proteins, the model protein, villin headpiece subdomain (HP35) consisting of 35 residues [43], was used to compute the disruption of proteins caused by PO with a size of 38.1 Å × 43.1 Å. Due to its small size and fast folding dynamics, HP35 has been chosen as a model for many computational and experimental studies [44-48]. HP35 was prepared from the Protein Data Bank (PDB code: 1YRF [43]) and modeled with the Amber03 force field [49].

Based on the oxidation theory of phosphorene by Ziletti et al. [38], we constructed the PO nanosheets with three oxidation modes (see Figure S1 in Supporting Information): dangling oxidation (mode I), interstitial oxidation (mode II), and horizontal bridge oxidation (mode III) to interact with HP35. In order to investigate the effects of oxidation concentrations on the toxicity of PO nanosheets, we constructed PO nanosheets with three concentrations of 2%, 5%, and 10% for each oxidation mode. According to previous studies [42, 50, 51], phosphorus atoms of phosphorene were modeled as uncharged Lennard-Jones particles with a cross section $\sigma_{PP}$= 0.330 Å and a depth of potential well $\varepsilon_{PP}$= 0.400 $kcal\ mol^{-1}$. The oxygen atoms and phosphorus atoms connected to oxygen were modeled as charged Lennard-Jones particles. Based on the CASTEP module in Material Studio, we calculated the differential charge density and charge distribution of oxygen and phosphorus atoms of PO. The calculated charge amount is given in Table S1 in the Supporting Information.

We constructed a PO-HP35 complex and placed it in the center of a periodic cube box (see Figure S2 in Supporting Information). The initial distance between the protein and the geometric center of PO was set as 32.3 Å. The combined systems were solvated in a cubic box with the periodic boundary condition. The distance between the solutes and the box boundary is at least 10 Å. The solvent was modeled with the TIP3P water model, and two chloride ions were added into the solution to ensure that the whole system is electrically neutral.

The solvated systems were simulated with MD method and Gromacs package version 4.5.5 [52]. During the simulation, PO and HP35 are free to move in space and the cutoff for the van der Waals interaction was set to 10 Å. The long-rang electrostatic interactions were treated with the particle-mesh Ewald method [53], with a grid spacing of 1.2 Å. After energy minimization, all systems were equilibrated by MD simulation for at 200 ps at a constant pressure of 1 bar and temperature of 300 K using Berendsen coupling [54]. Then all simulations were performed in an NVT ensemble at 300 K. The VMD software was used to visualize and analyze the results of MD simulations [55].

## 3. RESULTS AND DISCUSSION

### 3.1. Effect of the oxidation mode of PO on the PO-HP35 interaction

The biological function of proteins is dependent on their natural structure. The biological toxicity of nanomaterials mainly comes from the destruction of the structure of proteins by the materials, causing them to lose their biological functions. In order to evaluate the effect of PO's oxidation mode on the PO-HP35 interaction, we first fixed the oxidation concentration of PO at 2% and then investigated the interaction between HP35 and PO with three oxidation modes at this oxidation concentration. Figure 1 (a) and (b) show the root mean square deviation (RMSD) and the number of α-Helix of the protein HP35 interacting with PO with the three oxidation modes, respectively. It is shown that under the three oxidation modes of PO, the RMSD of protein HP35 does not change much over time. The RMSD of HP35 in mode II and mode III of PO is about 0.1~0.2 nm, and it is about 0.2 nm on average in mode I. The RMSD value of HP35 in the three oxidation modes of PO at oxidation concentration 2% is close to that of HP35 in the absence of nanomaterials (see Figure S3 in Supporting Information). Thus, the effect of PO's oxidation mode on the structure of HP35 can be ignored and PO at this oxidation concentration has no destructive effect on the structure of HP35.

Structurally, HP35 is mainly composed of three α-helices. Figure 1 (b) shows that under the three oxidation modes, the number of residues in the α-helix structure of HP35 changes very similarly over time. The average numbers of residues in the α-helix structure of HP35 in mode I, II and III are 21, 21.6, 21.8, respectively, which again shows that the effect of oxidation mode of PO is negligible. In-depth secondary structure analyses of HP35 were further conducted with the DSSP program [56] under the three oxidation modes. Figures 1 (c) shows the secondary structure of HP35 within the time scale of 0-500 ns under the three oxidation modes. It is shown that the secondary structure of HP35 under the three oxidation modes is basically unchanged within this time scale.

In order to more clearly show the effect of the oxidation mode of the PO nanosheet on the structure of HP35, Figure 2 shows the snapshots of HP35 on the surface of PO under three oxidation modes at 500 ns. It can be seen that the structure of HP35 on the surface of the PO under the three oxidation modes is not damaged by PO, and the three α-helices are clearly visible. Figure 1 and 2 show that the oxidation mode of phosphorene has almost no effect on the PO-HP35 interaction, and the same conclusion can also be obtained by using PO under other oxidation concentrations (see Figure S4 in Supporting Information).

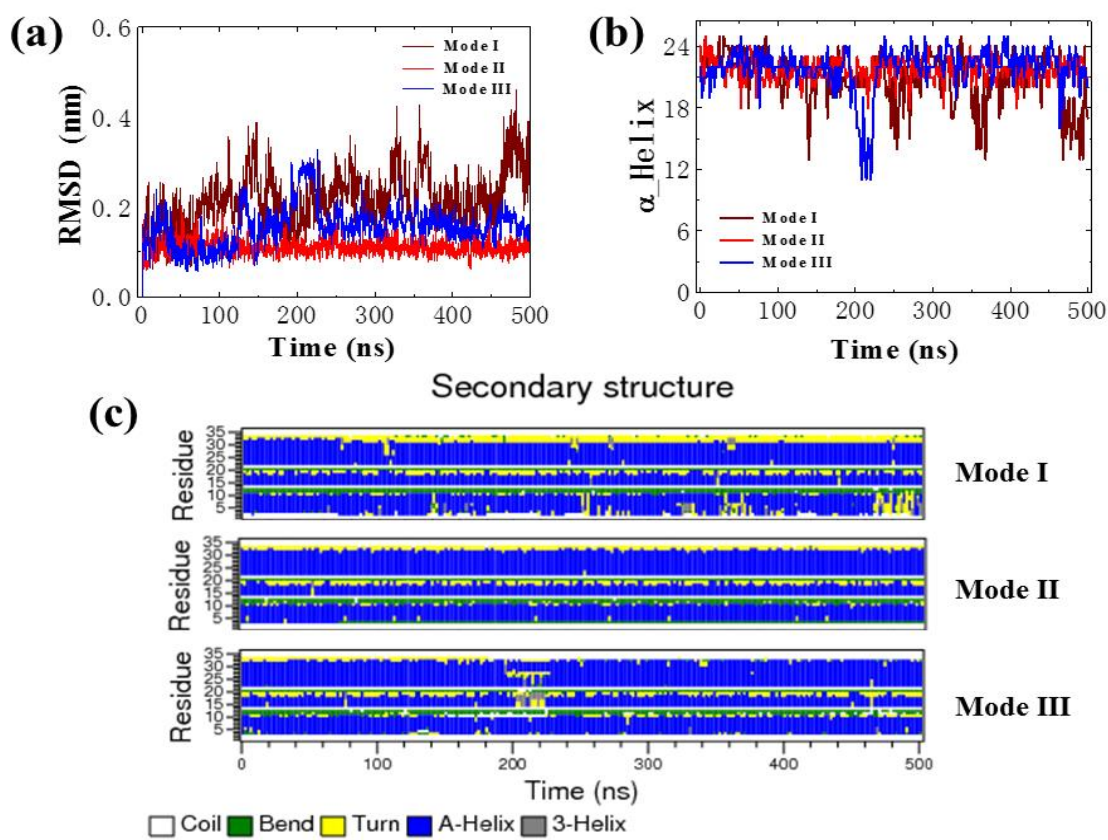

**Figure 1.** (a) The root mean square deviation (RMSD), (b) the number of residues in the α-helix structure and (c) the secondary structure of HP35 absorbed on the surface of PO nanosheets in oxidation mode I, II and III as a function of time. The oxidation concentration of PO is 2% for the three oxidation modes.

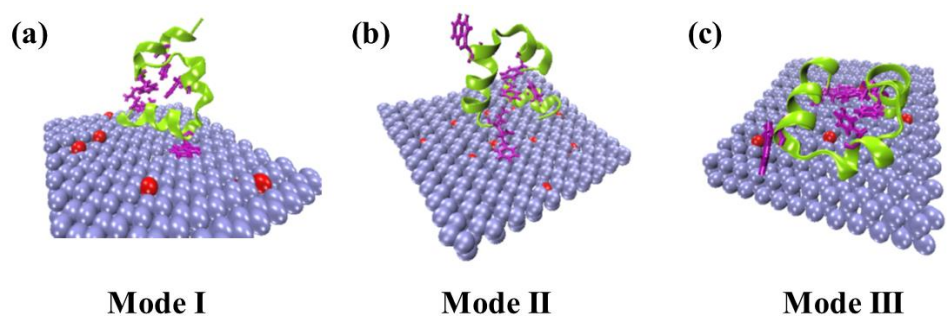

**Figure 2.** The snapshots of HP35 at 500 ns on the surface of PO with oxidation mode I (a), II (b) and III (c) at oxidation concentration 2%. For clarity, the water molecules are not shown.

## 3.2 Effect of oxidation concentration of PO on the PO-HP35 interaction

To investigate the effect of the oxidation concentration of PO on the interaction between PO and the HP35, we selected PO with three oxidation concentrations in oxidation mode II as a representative for analysis. There are similar situations in other oxidation modes of PO. Figure 3 (a) and (b) show the RMSD and the number of residues in the $a$-helix structure of HP35 under oxidation concentration 2%, 5% and 10% as a function of time, respectively. The RMSD of protein HP35 increases with the increase of oxidation concentration, as shown in Figure 3 (a). When the oxidation concentration of PO is 2%, the RMSD of protein HP35 is about 0.1-0.2 nm. In the case of 5% oxidation concentration, the RMSD of HP35 is about 0.3-0.4 nm in the time interval of 400-500 ns. While the oxidation concentration is 10%, the RMSD of HP35 is about 0.6-0.7 nm in the time interval of 400-500 ns. Figure 3 (b) shows that as the oxidation concentration of PO increases, the number of residues in the $a$-helix structure of HP35 decreases. In the time interval of 400-500 ns, the average number of residues in the $a$-helix structure of HP35 are 21.5, 17.5 and 5.1 for PO's oxidation concentration 2%, 5% and 10%, respectively.

Figure 3 (c) shows the effect of PO's oxidation concentration on the secondary structure of protein HP35. When the oxidation concentration of PO is 2%, the secondary structure of HP35 is basically unchanged in the time interval of 0-500 ns. When the oxidation concentration of PO is 5%, the first and third α-Helix chains have slight changes after 300ns, and part α-Helix turns into Turn and 3-Helix. When the oxidation concentration is 10%, all three α-Helix chains are destroyed, and especially the structures of the first and second α-Helix chains are almost completely destroyed after 40 ns.

In order to show more clearly the effect of the oxidation concentration of phosphorene on the structure of HP35, Figure 4 shows the snapshots of HP35 on the surface of PO with oxidation concentrations 2%, 5% and 10% at 500 ns. As shown in Figure 4, that when the oxidation concentration is 2%, the structural integrity of HP35 is hardly affected by PO; when the oxidation concentration is 5%, the structure of HP35 is slightly deformed, but most of the α-helix structure is still preserved; when the oxidation concentration is 10%, the structure of HP35 has changed dramatically, and most of the original three α helixes no longer exist. Figures 3 and 4 show that the greater the oxidation concentration, the greater the structural damage and toxicity of PO to HP35.

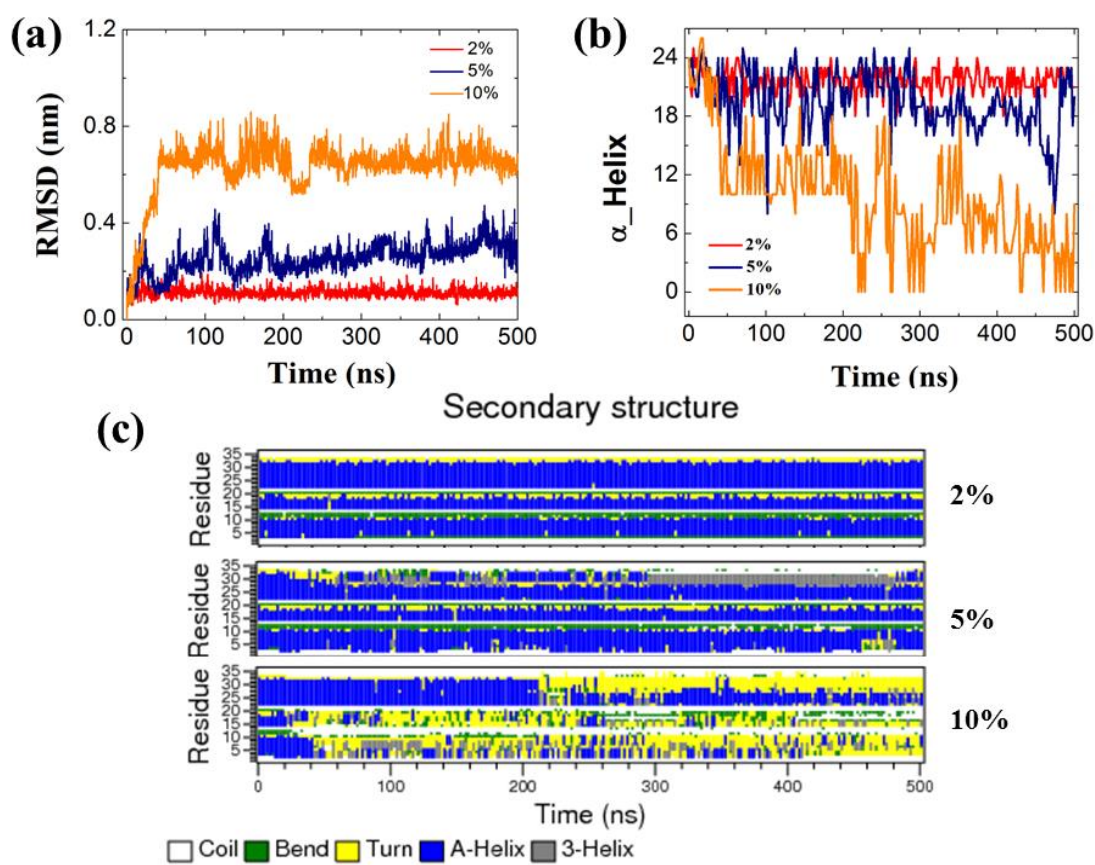

**Figure 3.** (a) The root mean square deviation (RMSD), (b) *a*-helical residue number and (c) protein secondary structure of protein HP35 absorbed on the surface of PO nanosheets with the oxidation concentration 2%, 5% and 10%. The oxidation modes of PO with the three oxidation

concentrations are Model II.

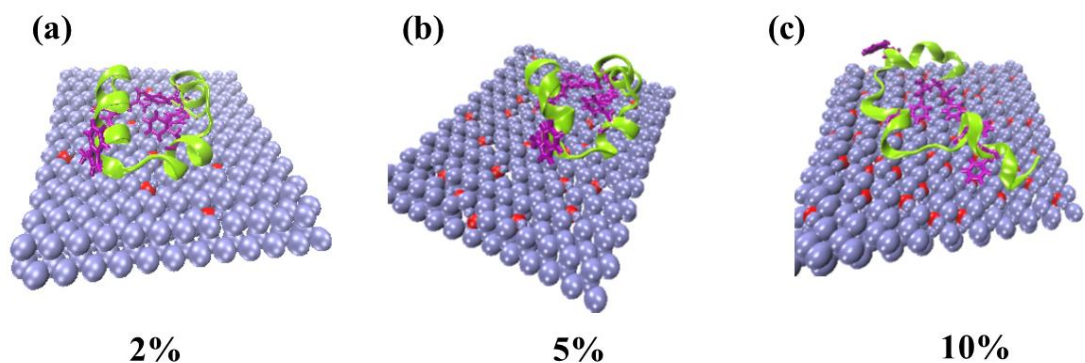

**Figure 4.** The snapshots of HP35 on the surface of PO with oxidation concentrations 2%, 5% and 10% at 500 ns.

In order to show the disruption process of PO with high oxidation concentration on the structure of HP35, Figure 5 shows the structural evolution of HP35 over time on the surface of PO with the oxidation concentration of 10% in oxidation mode II. During the interaction between HP35 and PO, the five aromatic amino acid residues play a major role in the structural changes of HP35. The five aromatic amino acid residues are: F06, F10, F17, W23 and F35, which are shown in bold with purple rods in the figure. Aromatic amino acid residues and PO form π-Σ interaction, which make the aromatic amino acid residues tightly adsorb on the surface of PO like an anchor, thereby tearing the protein structure. Figure 5 shows that HP35 was adsorbed to the surface of PO at 22 ns. Except W23 and the remaining aromatic amino acid residues: F06, F10, F17 and F35 were all adsorbed to the surface of PO at 28 ns, and the structure of the protein began to change. The protein structure changes significantly at 42 ns, but at this time the three α-helix structures are still complete. At 136 ns, the first and second α-helix chains have been destroyed, and the structure of the third helical chain is still intact; after 216 ns, all three α-helix structures have been destroyed.

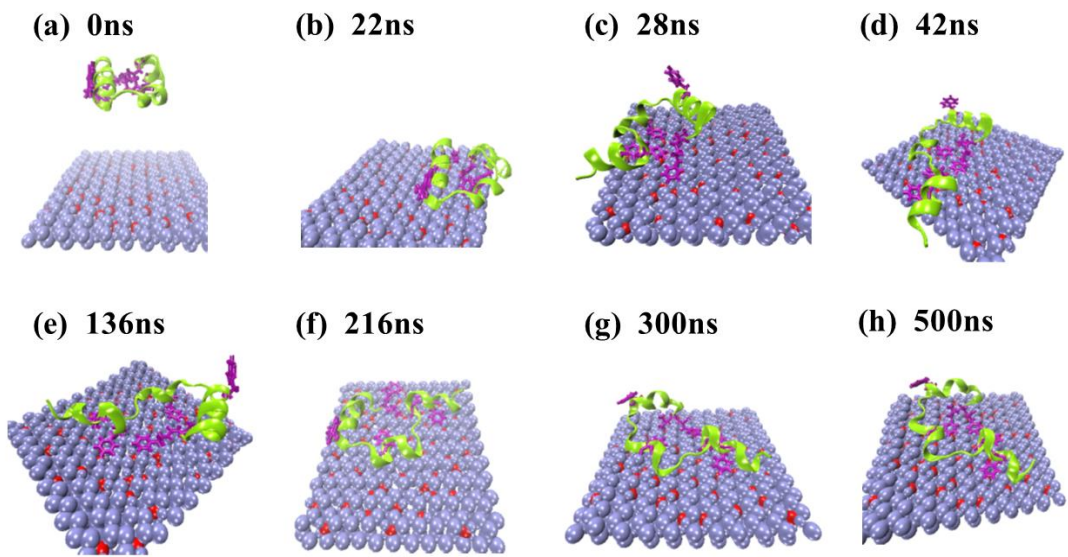

**Figure 5.** The snapshots of the evolution of HP35 on the surface of PO (mode II with the oxidation concentration 10%) at different times. The aromatic residues F06, F10, F17, W23 and F35 are shown as purple rod-like structure.

To gain more insight into the influences of PO nanosheets on HP35, we further examined the local structure of HP35 by calculating the two dihedral angles $\phi=C_{(-1)}NC_\alpha$ and $\Psi=NC_\alpha CN_{(+1)}$. Figure 6 (a), (b) and (c) show the Ramachandran plots of HP35 adsorbed onto PO nanosheets with oxidation concentration 2%, 5% and 10%, respectively. In Figure 6 (a), most of the population is concentrated in the regions (−75°, −40°) and (−75°, 150°), corresponding to the α-helix and the so called "polyproline II" (PPII) regions, respectively. With the increase of the oxidation concentration of PO, the distribution of population becomes more dispersed. Then, we calculated the probability distribution of each dihedral Angle of HP35 adsorbed onto PO with the three oxidation concentrations as shown in Figure 6 (d) and (e). As the oxidation concentration of PO increases, the probability distribution of φ of HP35 adsorbed onto PO decreases around the values −75° and 60°, but increases around the values −120° and 120°. The probability distribution of Ψ of HP35 adsorbed onto PO increases in the regions of −180° to −120° and 60° to 120°, but decreases around the value −40° with the increase of the oxidation concentration of PO. Compared with the other two cases, the probability of Ψ of HP35 adsorbed onto PO with the oxidation concentration 10% is smallest around the value 150°.

In short, the data of the RMSD, *a*-helical residue number of HP35, the mapping of its secondary structure changes and the profiling of its dihedral angles reveal that oxidation enhances the damage of phosphorene to the structure of HP35. The greater the oxidation concentration, the stronger the disruption of the structure of HP35 caused by PO.

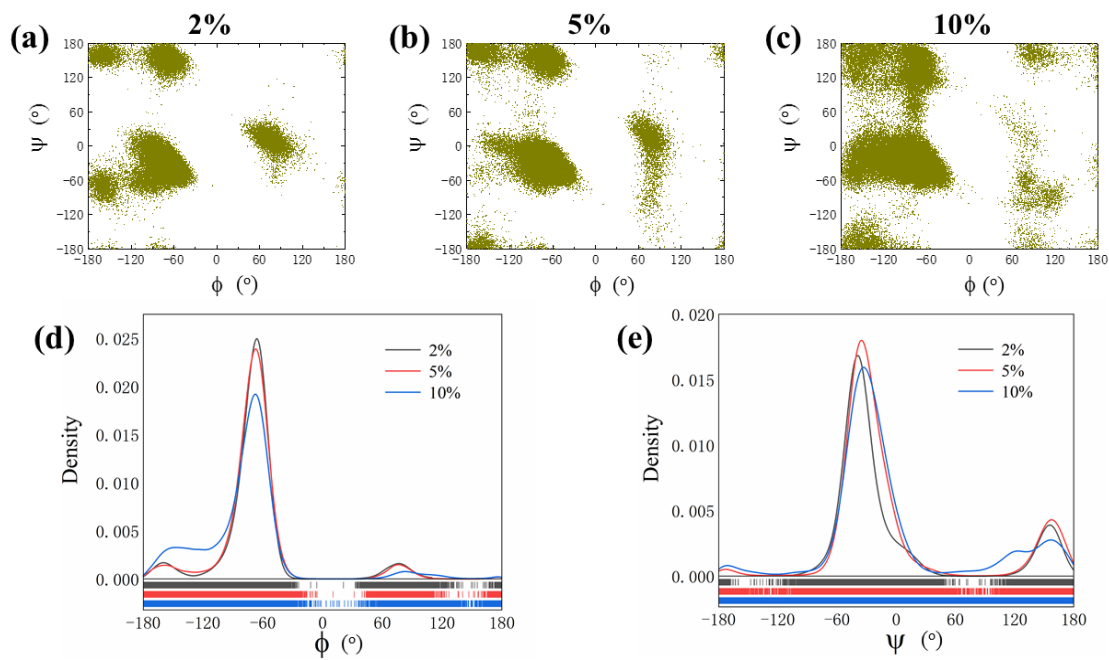

**Figure 6.** (a, b and c) Ramachandran plot of HP35 adsorbed onto PO (mode II) with the oxidation concentration 2%, 5% and 10%. (d and e) Distribution of the dihedral angles ϕ and Ψ of HP35 adsorbed onto PO with the three oxidation concentrations.

In order to understand the effect of phosphorene oxidation on the structure of HP35, Figure 7 shows the interaction energy $E$, $E_{vdw}$ and $E_{col}$ between HP35 and PO nanosheets as a function of time, where $E = E_{vdw} + E_{col}$, $E_{vdw}$ and $E_{col}$ are Van der Waals and Coulomb interaction energy, respectively. At time t = 0 ns, the interaction energy $E$ for all three oxidation concentrations is zero, as shown in Figure 7 (a), indicating that HP35 and PO nanosheets were separated initially. At about t = 70 ns, the interaction energy $E$ between HP35 and PO nanosheets with the oxidation concentration of 2% and 5% is essentially stable. However, the interaction energy $E$ between HP35 and PO with an oxidation concentration of 10% does not reach stability until about 430 ns, and $E$ increases with time before this time. After the interaction energy $E$ is stabilized, the greater the oxidation concentration, the greater the absolute value of $E$, as shown in Figure 7 (a). The interaction energy $E$ consists of two parts:

Coulomb interaction energy $E_{col}$ and van der Waals interaction energy $E_{vdw}$. The atoms that have a Coulomb interaction with HP35 are the oxygen atom of PO and the phosphorus atom connected to the oxygen atom. Compared with the Coulomb interaction energy, the van der Waals interaction energy between the oxygen atom of PO and HP35 is stronger, as shown in Figure 7 (b). The Coulomb interaction energy difference between HP35 and oxygen atom and phosphorus atom is very small (see Figure S5 in Supporting Information). The Coulomb interaction energy $E_{col}$ of PO with different oxidation concentrations and HP35 are not much different, as shown in Figure 7 (c). However, the van der Waals interaction energy between the oxygen atoms of PO with different oxidation concentrations and HP35 is significantly different, as shown in Figure 7 (d). The greater the oxidation concentration, the stronger the van der Waals interaction energy between the oxygen atoms of PO and HP35. Van der Waals interaction energy $E_{vdw}$ makes a major contribution to the total energy $E$. After oxidation, oxygen atoms fill the groove region in puckered surface of phosphorene (see Figure S1 in Supporting Information), which enhances the dispersion effect of phosphorene. This is the main reason why PO nanosheets with higher oxidation concentration has stronger disruption to the structure of HP35.

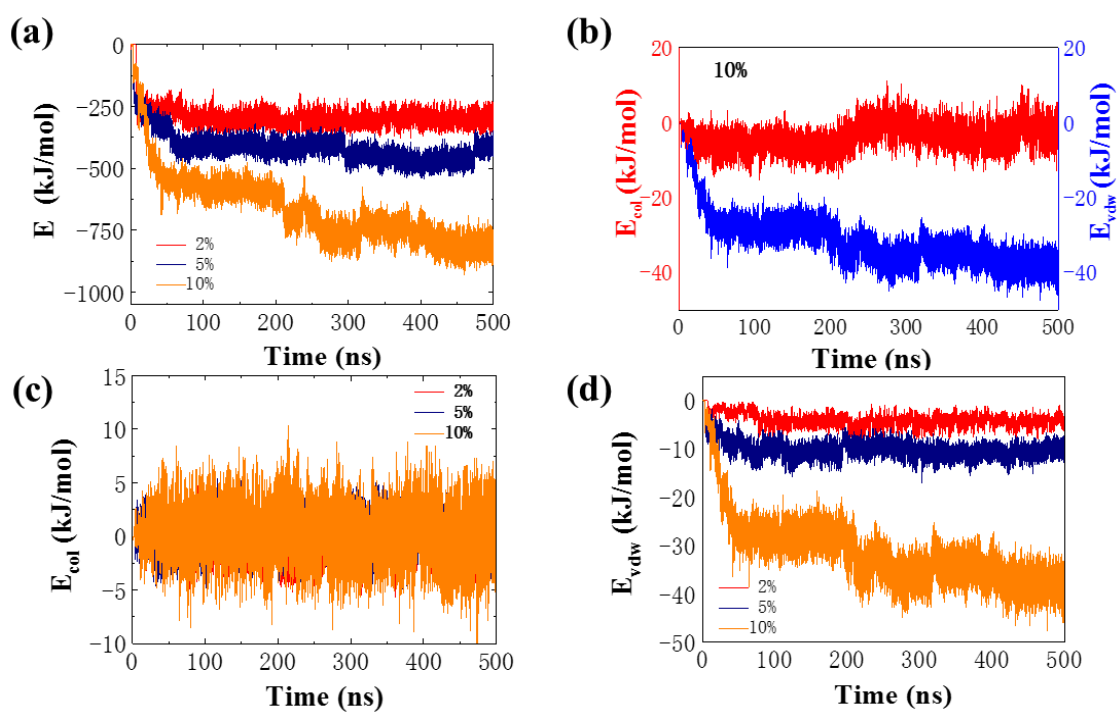

**Figure 7.** (a) The interaction energy $E$ between HP35 and PO (in Mode II with the oxidation concentration of 2%, 5% and 10%) as a function of time. (b) Coulomb interaction energy $E_{col}$ and Van

der Waals interaction energy $E_{vdw}$ between HP35 and the oxygen atom of PO (in Mode II with an oxidation concentration of 10%) as a function of time. (c) Coulomb interaction energy $E_{col}$ between HP35 and PO (in Mode II with the oxidation concentration of 2%, 5% and 10%) as a function of time. (d) van der Waals interaction energy $E_{vdw}$ between HP35 and oxygen atom of PO (in Mode II with the oxidation concentration of 2%, 5% and 10%) as a function of time.

## 4. Conclusions

Phosphorene is a new member of the two-dimensional material family and has great potential for future applications, especially in the fields of biomedicine and biosensors. However, phosphorene is unstable in an aerobic environment and is easily oxidized to form phosphorene oxide (PO). It is necessary to investigate PO's interactions with proteins, and evaluate its biological responses. In this paper, using large scale molecular dynamics simulation, we investigated the interactions between the model protein HP35 and the PO nanosheets in oxidation mode I, II, and III with oxidation concentrations of 2%, 5%, and 10%. The simulation results demonstrated that the oxidation mode of PO nanosheets has little effect on the PO-HP35 interaction, but the oxidation concentration has a great influence on it. The greater the oxidation concentration of PO, the stronger the disruption of the PO nanosheets to the structure of HP35. The van der Waals interaction between HP35 and PO with higher oxidation concentration is stronger, which makes oxidation enhance the disruption of phosphorene nanosheet to the structure of model protein HP35. The oxygen atoms filled into the groove region of the puckered surface increase the dispersion interaction between phosphorene and HP35. The disruption process of PO with high oxidation concentration on the structure of HP35 and the details of the structural changes of HP35 were also presented in the paper.

The results in this paper would help understanding phosphorene's biological toxicity and might shed light on designing the phosphorene-protein interactions. Preventing phosphorene oxidation and reducing its biological toxicity would be an important direction in the future. These results should also provide insight for a better understanding of the interactions between proteins and other puckered monolayer materials such as MoS2 and WS2.

# AUTHOR INFORMATION

## Corresponding Author

*E-mail: wzhangph@cumt.edu.cn (W. Z.); yangxianqing@cumt.edu.cn (X. Y.)

## ORCID

Wei Zhang: 0000-0001-5423-1240

## Notes
The authors declare no competing financial interests.

# Acknowledgements

This work was partially supported by the National Natural Science Foundation of China (Grant No. 11774417) and the Basic Research Program Project of Xuzhou (Grant No. KC21020).

# Supporting Information

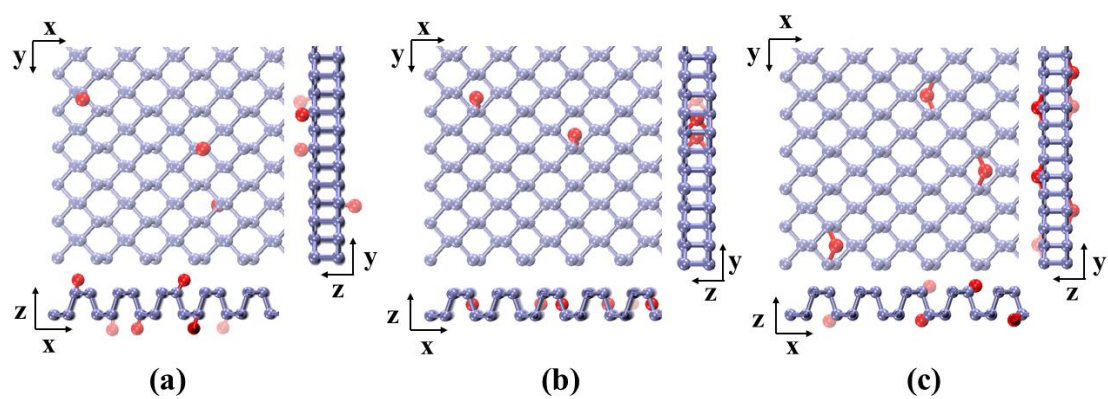

**Figure** S1. Oxidation mode of phosphorene nanosheet. (a) Dangling oxidation (Mode I). (b) Interstitial oxidation (Mode II). (c) Horizontal bridge oxidation (Mode III).

**Table S1. Charge Distribution in PO nanosheets**

| Oxidation Mode | Phosphorus/e | Oxygen/e |
|---|---|---|
| **Dangling oxidation** | 0.21 | -0.21 |
| **Interstitial oxidation** | 0.08 | -0.16 |
| **Horizontal bridge oxidation** | 0.09 | -0.18 |

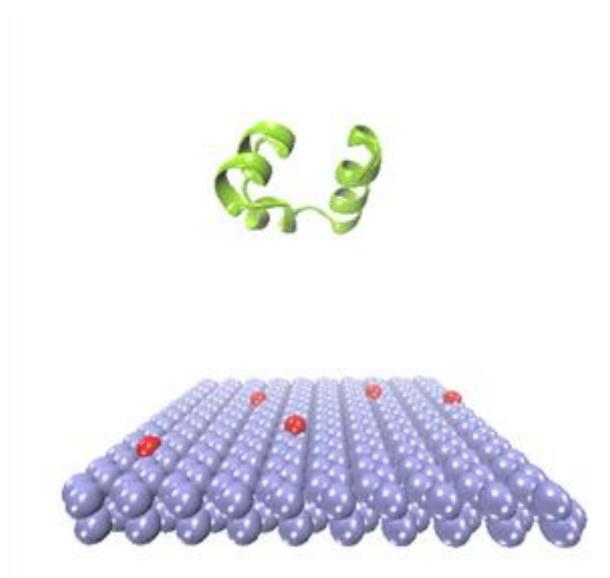

**Figure S2**. A complex composed of phosphorene oxide (Mode I) and HP35. The red, blue and yellow spheres represent oxygen atoms, phosphorus atoms and HP35, respectively.

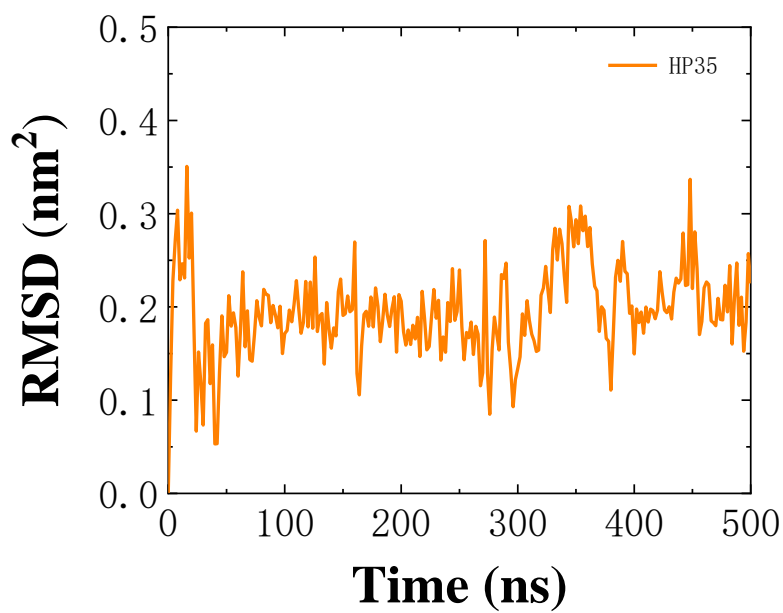

**Figure S3.** The root mean square deviation (RMSD) of protein HP35 in the absence of nanomaterials.

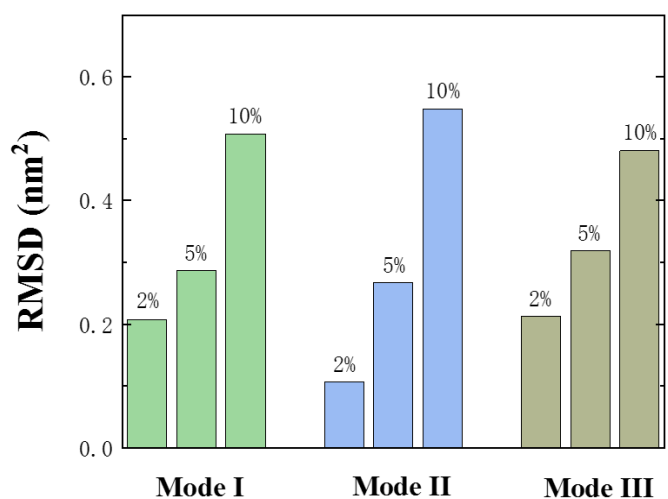

**Figure S4.** The root mean square deviation (RMSD) of protein HP35 adsorbed on the surface of PO nanosheets in Mode I, II and III with the oxidation concentration of 2%, 5% and 10% at 200 ns.

As shown in Figure S4, the oxidation mode of PO has almost no effect on the RMSD of HP35 for a certain oxidation concentration of PO. However, the oxidation concentration of PO has a significant effect on the RMSD of HP35. The greater the oxidation concentration of PO, the greater the RMSD of HP35, indicating that the disruption of PO to HP35 increases as the oxidation concentration increases. In the three oxidation modes of PO, the RMSD of HP35 varies with the oxidation concentration in the same trend.

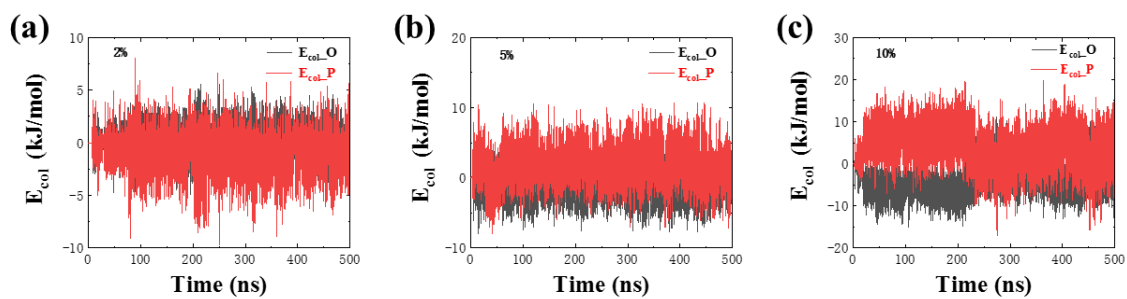

**Figure S5.** The Coulomb interaction energy between HP35 and oxygen atom and phosphorus atom of PO with Mode II and the oxidation concentration of (a) 2%, (b) 5% and (c) 10%.